\documentstyle[12pt,worldsci]{article}

\def\PL #1 #2 #3 {Phys. Lett.~{\bf#1} (#2) #3}
\def\NP #1 #2 #3 {Nucl. Phys.~{\bf#1} (#2) #3}
\def\ZP #1 #2 #3 {Z.~Phys.~{\bf#1} (#2) #3}
\def\PR #1 #2 #3 {Phys. Rev.~{\bf#1} (#2) #3}
\def\PRD #1 #2 #3 {Phys. Rev.~D {\bf#1} (#2) #3}
\def\PP #1 #2 #3 {Phys. Rep.~{\bf#1} (#2) #3}
\def\PRL #1 #2 #3 {Phys. Rev.~Lett.~{\bf#1} (#2) #3}

\newcounter{eqletter}

\begin{document}
\baselineskip14pt
\title{\mbox{GLUON RADIATION PATTERNS IN HARD SCATTERING EVENTS}\footnote{Talk
presented at the Conference on Physics {\it Beyond the
Standard Model IV}, Lake Tahoe, California, December 14--18, 1994.}
}
\author{\small D.~Zeppenfeld\\
{\it Department of Physics, University of Wisconsin, 1150 University Ave.\\
Madison, WI 53706, USA}\\
E-mail: dieter@phenom.physics.wisc.edu}
\maketitle

\begin{center} ABSTRACT\\ [.1in]
\parbox{13.5cm}{\small
The radiation pattern of relatively soft gluons in hard scattering events
is sensitive to the underlying color structure. As an example
I consider heavy Higgs production via weak boson fusion at the LHC.
A minijet veto, which makes use of the different patterns for signal and
backgrounds, provides an effective Higgs search tool.   }
\end{center}


Finding ways to detect a heavy Higgs boson or longitudinal
weak boson scattering at the LHC is an issue of highest importance as
long as the nature of spontaneous electroweak symmetry breaking remains to
be established. In order to distinguish weak boson scattering, $i.e.$ the
electroweak process $qq\to qqVV$, from large backgrounds due to QCD
processes and/or the production of $W$ bosons from the decay of top quarks,
tagging of at least one fast forward jet is essential~\cite{Cahn}. Early
studies~\cite{Froid} showed that double tagging is quite
costly to the signal rate because one of the two quark jets has substantially
lower median $p_T$ (order 30 GeV) than the other (order 80 GeV). Single
forward jet tagging relies only on the higher $p_T$ tag-jet and thus proves
an effective technique~\cite{BCHZ,DGOV}.

A study of the $WW$ signal must exploit additional identifying
characteristics. For example, the $W$ bosons from top quark decays can
be rejected by vetoing the additional central $b$ quark jets arising in
$t\to Wb$~\cite{BCHZ}. For 
$H\to W^+W^- \to \ell^+\nu\ell^-\bar\nu$ another important discriminator is
a large transverse momentum difference between the charged
leptons~\cite{DGOV}.

In a weak boson scattering event no color is exchanged between
the initial state quarks. Color coherence between initial and final state
gluon bremsstrahlung then leads to a suppression of hadron production in the
central region, between the two tagging jet candidates of the
signal~\cite{troyan,Kane}. Typical backgrounds like $t\bar t$
production or QCD jet emission in $W^+W^-$ production involve color exchange
between the incident partons and, as a result, gluon radiation into the
central region dominates.

A second distinction is the momentum scale of
the hard process which governs additional gluon radiation. In longitudinal
weak boson scattering the color charges, carried by the incident quarks,
receive a momentum transfer of order the transverse momentum of the final
state quarks which typically is in the $Q=30$ to 80~GeV range.
For the background processes, on the other hand, the color charges receive
a much larger momentum kick, of the order of the weak boson pair mass or even
the parton center of mass energy of the event, {\it i.e.} $Q\approx 1$~TeV.
Extra parton emission is suppressed by a factor
$f_s=\alpha_s {\rm ln}\; (Q^2/p_{T,{\rm min}}^2)$, where $p_{T,{\rm min}}$ is
the minimal transverse momentum required for a parton to qualify as a jet.
The jet transverse momentum scale below which multiple minijet emission must
be expected is set by $f_s=1$. Because the hard scale $Q$ is much larger
for the backgrounds than for the signal a veto on additional
minijet activity should provide an efficient tool to suppress the backgrounds,
with little cost to the signal. In effect such a technique constitutes a
rapidity gap trigger at the minijet level instead of the soft hadron level
as was suggested previously~\cite{troyan}. At the LHC the low
signal cross sections require running at high luminosity and then overlapping
events in a single bunch crossing will likely fill a rapidity gap with
soft hadrons even if it is present at the level of a single $pp$ collision.
However, even at ${\cal L} = 10^{34}{\rm cm}^{-2}{\rm sec}^{-1}$ only about
20\% of random bunch crossings are expected to lead to a jet with
$p_T > 20$~GeV~\cite{ciapetta} and hence a rapidity gap of minijets may well
be observable at design luminosity.

In a recent paper~\cite{bpz} these ideas were analyzed in detail for a
particular example of longitudinal weak boson scattering: production and
subsequent decay $H\to W^+W^- \to \ell^+\nu\ell^-\bar\nu$ of a heavy
Higgs boson. 
I would like to summarize the results
in the remainder of this talk. At the same time it should be stressed that
the basic method is more general: it can be applied to the study of
longitudinal weak boson scattering in all channels.

The analysis of Ref.~8 is based on full tree level simulations of the
partonic subprocesses for the signal and the backgrounds. The backgrounds
considered are $q\bar q \to W^+W^-$ with additional QCD radiation of up to two
partons, $pp \to t\bar t +{\rm n\; jets},\; {\rm n}=0,1,2$ with
subsequent top quark decays $t\bar t \to bW^+\bar b W^-$, and
the electroweak background from transversely polarized $W$'s in weak boson
scattering subprocesses like $qq\to qq(g)W^+W^-$. This electroweak
background is taken as the SM cross section without a heavy Higgs boson
and thus the signal is defined as $\sigma(m_H)-\sigma(m_H=100\; {\rm GeV})$.
In all cases the leptonic decays of the $W$'s are implemented in the narrow
width approximation, at the amplitude level.

In order to identify the signal we first consider events with two well
isolated, central leptons ($\ell=e,\mu$):
\begin{eqnarray}\label{cut1}
p_{T\ell}  >  50\; {\rm GeV}\;, \qquad |\eta_\ell|  <  2 \; &,& \qquad
R_{\ell j} = \sqrt{(\eta_\ell-\eta_j)^2 + (\phi_\ell-\phi_j)^2}  >  0.7\;,
\nonumber \\
\Delta p_{T\ell\ell} = | {\bf p}_{T\ell_1}-{\bf p}_{T\ell_2}| & > & 300\,
{\rm GeV}\;, \qquad m_{\ell\ell}> 200\, GeV\;.
\end{eqnarray}
The $R_{\ell j}>0.7$ separation cut forbids a parton
(jet) of $p_T>20$~GeV in a cone of radius 0.7 around the lepton direction.
The cut on $\Delta p_{T\ell\ell}$, the difference of the charged lepton
transverse momentum vectors~\cite{DGOV}, is crucial to concentrate on high
invariant mass and high $p_T$ $W$ pairs. The cross sections after these
lepton cuts are given in the first column of the table.

Next we require the existence of a tagging jet, which is taken as the highest
transverse momentum parton, which must then satisfy
\begin{equation}\label{cut2}
p_{Tj}^{\rm tag} > 50\, {\rm GeV}\;, \qquad E_j^{\rm tag} > 500\, {\rm GeV}\;,
\qquad 1.5 <|\eta_j^{\rm tag}| < 4.5 \;.
\end{equation}
In addition the tagging jet must be well separated from the $W$ decay leptons,
\begin{equation}\label{cut3}
{\rm min}\; |\eta_j^{\rm tag}-\eta_\ell | > 1.7\; ,
\end{equation}
The signal and background cross sections after the cuts of Eqs.~\ref{cut2}
and \ref{cut3} are listed in the second and third columns of the table,
respectively.

\begin{table}
\caption{Signal and background cross sections $B\sigma$ in fb after
increasingly stringent cuts. Four leptonic decay channels of the $W^+W^-$ pair
are included. 
}
\vglue0.1in
\tabcolsep=.35em
\begin{tabular}{lcccc}
& lepton cuts only& + tagging jet&
{\def\arraystretch{.66}
\begin{tabular}[t]{c}
+ lepton-\\
tagging jet\\
separation
\end{tabular}}
& {\def\arraystretch{.66}
\begin{tabular}[t]{c}
+ minijet veto\\
($p_{T,\rm veto}=$\\
20~GeV)
\end{tabular}}\\
& [Eq.~(\ref{cut1})]& [Eq.~(\ref{cut2})]& [Eq.~(\ref{cut3})]&
[Eq.~(\ref{cutveto})]\\
\hline
$WW(jj)$& 27.4& 1.73& 0.57& 0.13\\
$t\bar t(jj)$& 640& 57& 25& 0.47\\
$m_H=100$ GeV& 1.18& 0.56& 0.29& 0.18\\
$m_H=800$ GeV& 3.4& 1.79& 1.31& 0.97\\
signal& 2.2& 1.23& 1.02& 0.79
\end{tabular}
\end{table}
\vglue0.2in

\begin{center}
\vglue2.9in
\includegraphics{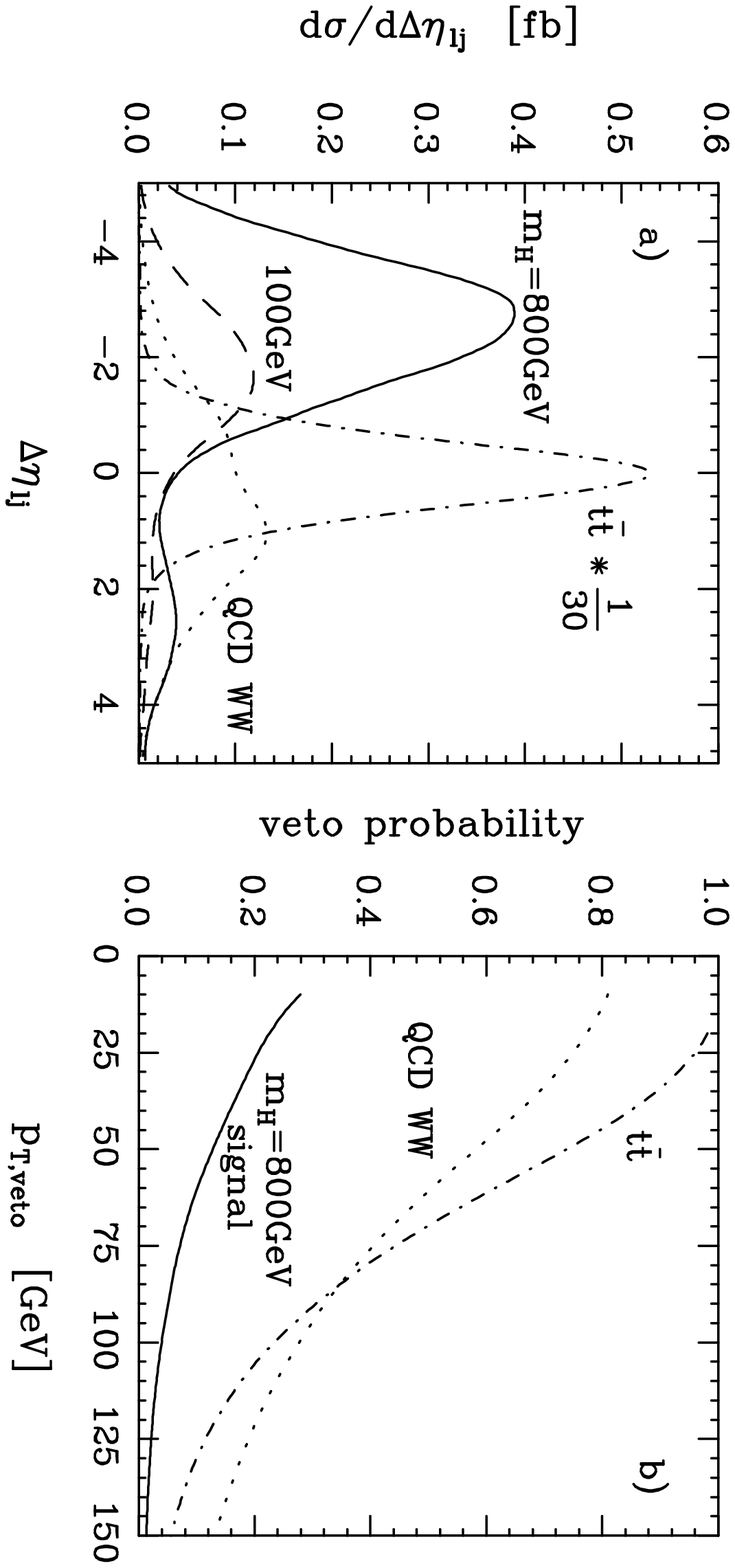}

\parbox{13.5cm}{\small\baselineskip12pt
 Fig.~1: Rapidity and transverse momentum distributions of secondary jets.
In a) $\Delta\eta_{\ell j}$ measures the pseudorapidity distance of the jet
closest to the leptons from the average lepton rapidity $\bar\eta$.
Negative values of $\Delta\eta_{\ell j}$ correspond to soft jets on the
opposite side of the leptons with respect to the tagging jet. The dashed
line shows the distribution for the electroweak background as defined by the
$m_H=100$~GeV case. The $t\bar tjj$ background has been scaled
down by a factor $1/30$. The probability to find a veto jet candidate
above a transverse momentum $p_{T,\rm veto}$ in the veto region of
Eq.~(\ref{cutveto}) is shown in b).
}
\end{center}

The cuts discussed so far define the hard scattering event. What are the
features of the soft radiation patterns in these events? Fig.~1a shows the
angular distribution of the jet (parton with $p_T> 20$~GeV) closest to the
leptons (more precisely, closest to the average lepton rapidity $\bar\eta =
(\eta_{\ell^+}+\eta_{\ell^-})/2$). The background processes favor emission
close to the leptons while the closest jet in the Higgs signal is typically
the second quark in the $qq\to qqH$ process. Vetoing any jets in the veto
region defined by
\begin{equation}\label{cutveto}
p_{Tj}^{\rm veto} > p_{T,\rm veto}\;, \qquad
\qquad \eta_j^{\rm veto} \varepsilon \;\;
[\eta_\ell^{\rm min}-1.7,\eta_j^{\rm tag}]\;\; {\rm or} \;\;
[\eta_j^{\rm tag},\eta_\ell^{\rm max}+1.7]\; ,
\end{equation}
will thus substantially reduce the backgrounds while having little effect
on the Higgs signal. The veto probability as a function of the cut value
$p_{T,\rm veto}$ is shown in Fig.~1b. Even though these results were obtained
in the truncated shower approximation and a more precise modeling is needed,
they clearly demonstrate that, in the central region, the backgrounds have a
much higher probability to produce additional minijets from QCD radiation.
Even in the $t\bar t$ background, where a strong suppression is obtained by
vetoing the central $b$-quark jets arising from the top decays, the veto on
the jet activity from soft QCD radiation provides an additional
background suppression by a factor 2 for $p_{T,\rm veto}=20$~GeV. Final cross
section values for signal and backgrounds are given in the last column of the
Table.

Should minijet vetoing be possible at the LHC for even smaller
$p_{T,\rm veto}$ values
the $t\bar t$ background to $H\to WW$ events can effectively be eliminated.
If low transverse momentum jets ($p_T \approx 30$~GeV) can also be
identified in the forward region, then double forward jet-tagging will provide
an additional strong suppression of the top quark background.

\section*{Acknowledgements}
I would like to thank V.~Barger and R.~Phillips for the most enjoyable
collaboration which lead to the results summarized in this talk.
This research was supported in part by the University of Wisconsin Research
Committee with funds granted by the Wisconsin Alumni Research Foundation,
by the U.~S.~Department of Energy under Grant No.~DE-FG02-95ER40896.



\begin{thebibliography}{99}
%
\bibitem{Cahn}
R.~N.~Cahn {\it et al.}, Phys.\ Rev.\ {\bf D35}, 1626 (1987);
V.~Barger {\it et al.}, 
Phys.\ Rev.\ {\bf D37} 2005 (1988);
R.~Kleiss and W.~J.~Stirling, Phys.\ Lett.\ {\bf 200B}, 193 (1988).

\bibitem{Froid}
U.~Baur and E.~W.~N.~Glover, Nucl.\ Phys.\ {\bf B347}, 12 (1990);
D.~Froideveaux, in {\it Proceedings of the ECFA Large Hadron
Collider Workshop}, Aachen, Germany, 1990, edited by G.~Jarlskog and D.~Rein
(CERN report 90-10, Geneva, Switzerland, 1990), Vol~II, p.~444;
M.~H.~Seymour, {\it ibid}, p.~557.

\bibitem{BCHZ}V.~Barger {\it et al.}, 
Phys.\ Rev.\ {\bf D44}, 2701 (1991); {\bf 48} 5444E (1993);
{\bf 48}, 5433 (1993).

\bibitem{DGOV}
D.~Dicus {\it et al.}, 
Phys.\ Lett.\ {\bf B258}, 475 (1991);
Nucl. \ Phys.\ {\bf B377}, 31 (1991).

\bibitem{troyan}
Y.~L.~Dokshitzer, V.~A.~Khoze, and S.~Troyan, in {\it
Proceedings of the 6th International Conference on Physics in Collisions},
(1986) ed.\ M.~Derrick (World Scientific, Singapore, 1987) p.365;
J.~D.~Bjorken, Int.\ J.\ Mod.\ Phys.\ {\bf A7}, 4189 (1992);
Phys.\ Rev.\ {\bf D47}, 101 (1993); 
preprint SLAC-PUB-5823 (1992).

\bibitem{Kane}
J.~F.~Gunion {\it et al.}, {\it Phys.~Rev.}\ {\bf D40}  (1989) 2223.

\bibitem{ciapetta}
G.~Ciapetta and A.~DiCiaccio, in {\it Proceedings of the ECFA Large Hadron
Collider Workshop}, Aachen, Germany, 1990, edited by G.~Jarlskog and D.~Rein
(CERN report 90-10, Geneva, Switzerland, 1990), Vol~II, p.~155.

\bibitem{bpz}
V.~Barger, R.~J.~N.~Phillips, and D.~Zeppenfeld,
Phys.\ Lett.\ {\bf B346}, 106 (1995).


\end{thebibliography}
\end{document}